\documentclass[a4paper,preprint]{revtex4}
\usepackage[dvips]{graphicx}
\usepackage{amssymb,amsfonts,amsmath}
\usepackage{color}
\usepackage{bm}
\makeatletter

\begin{document}
\title{State Concentration Exponent as a Measure of Quickness in Kauffman-type Networks}
\author{Shun-ichi Amari$\mbox{}^1$, Hiroyasu Ando$\mbox{}^1$, Taro
Toyoizumi$\mbox{}^1$ and Naoki Masuda$\mbox{}^{2,3}$\\ $\mbox{}^1$RIKEN
Brain Science Institute, Hirosawa 2-1, Wako, Saitama 351-0198, Japan\\
$\mbox{}^2$Department of Mathematical Informatics, The University of
Tokyo, 7-3-1 Hongo, Bunkyo, Tokyo 113-8656, Japan\\ $\mbox{}^3$ PRESTO,
Japan Science and Technology Agency, 4-1-8 Honcho, Kawaguchi, Saitama
332-0012, Japan }

\begin{abstract}
We study the dynamics of randomly connected networks composed of binary Boolean elements and those composed of binary majority vote elements. We elucidate their differences in both sparsely and densely connected cases. The quickness of large network dynamics is usually quantified by the length of transient paths, an analytically intractable measure. For discrete-time dynamics of networks of binary elements, we address this dilemma with an alternative unified framework by using a concept termed state concentration, defined as the exponent of the average number of $t$-step ancestors in state transition graphs. The state transition graph is defined by nodes corresponding to network states and directed links corresponding to transitions. Using this exponent, we interrogate the dynamics of random Boolean and majority vote networks. We find that extremely sparse Boolean networks and majority vote networks with arbitrary density achieve quickness, owing in part to long-tailed in-degree distributions. As a corollary, only relatively dense majority vote networks can achieve both quickness and robustness.
\end{abstract}
\maketitle

\section{Introduction}

Networks of binary elements are useful tools for investigating a
plethora of dynamical behavior and information processing in
biological and social systems.  For example, various models of
associative memory are used to study neural information processing
\cite{Amari1972IEEE,Hopfield1982PNAS,Hertz1991book}.  Random Boolean
networks, also known as Kauffman nets, show rich dynamics and are
used to model gene regulation
\cite{Kauffman1969JTB,Kauffman1984PhysicaD,Kauffman1993book,bornholdt2008}.  Random
majority vote networks are often used to understand mechanisms for
ordering in neural information processing
\cite{Hertz1991book,amari1974,AYK1977,Amari1990IEEE}, gene regulation
\cite{bornholdt2008}, and collective opinion formation in social
systems \cite{Castellano2009RMP}.
We study the dynamics of such networks by using a simple
generative model of randomly connected Boolean and majority vote
elements in the cases of sparse and dense connectivity.

Properties desirable for the dynamics of networks of such binary units
include robustness and quickness. A system is defined to be robust when
the flipping of a small number of units' states does not eventually alter
the behavior of the entire network. For random
Boolean networks, the robustness has been quantified in
the context of damage spreading in cellular automata
\cite{Luque1997PRE,Luque2000PhysicaA,Kauffman2004PNAS,Shmulevich2004PRL}.

Dynamics is usually called quick if an orbit starting from an arbitrary
state reaches the corresponding attractor within a small number of steps
on average, i.e., with a short transient length of the dynamics.
However, even the average transient length, which apparently seems to be the most basic quantity to characterize the statistics of the transient length, may be difficult to evaluate because the transient length of the random Boolean networks seems analytically intractable and it obeys long-tailed distributions \cite{BL1996}.
Therefore, in this paper,
we theoretically study the quickness of dynamics by use of
a concept of state concentration instead of examining the transient length. To this end, we extend the previous statistical dynamical framework
\cite{amari1974,AYK1977,Amari1990IEEE}.
In particular, the exponent of concentration, which we introduce later,
is an analytically tractable quantity for measuring the quickness of
dynamics in random Boolean and majority vote networks. Using this
exponent, we investigate the compatibility of the robustness and
quickness in these two types of networks
in two cases of connectivity, i.e., sparse and dense connectivity.

For this purpose, we distinguish densely connected Boolean networks
(DBNs), sparsely connected Boolean networks (SBNs), densely connected
majority vote networks (DMNs), and sparsely connected majority vote
networks (SMNs). 
We elucidate the differences between the four dynamics. In particular, we
show that strong state concentration, accompanied by
a power law type of in-degree distribution with an exponential cutoff, occurs in
the majority vote networks (DMNs and SMNs) but not for the Boolean
networks except for extremely sparse cases.
Then, we argue that DMNs are the only type among
the four types of network that realizes both robustness and quickness.

\section{Model}

Let us consider a network of $n$ binary units. We define the discrete-time
dynamics of the network by
\begin{equation}
x_i(t+1)=f_i \left(x_1(t), \ldots,
  x_n(t)\right)\quad (1\le i\le n),
\end{equation}
 where $x_i(t)\in \left\{1, -1\right\}$ is the binary state of the $i$th
 unit at time $t$. For a random Boolean network, each $f_i$ is randomly and independently chosen from
 the $2^{2^n}$ Boolean functions on the $n$ units.
For a majority vote network,
\begin{equation}
f_i(x_1, \ldots, x_n) = \mbox{sgn} \left(\sum_{j=1}^n
  w_{ij}x_j \right) \; (1\le i\le n),
\end{equation}
where sgn indicates the sign function.  We consider an ensemble of
randomly generated majority vote networks where $w_{ij}$ are
independently and identically distributed Gaussian random variables. In
general, a constant or random threshold could be included in the
above dynamical expression, which we omit here for simplicity.  If the
value of $f_i \left(x_1, \ldots , x_n \right)$ depends only on randomly
chosen $K$ units for each $i$, the model is called the $K$-sparse
network \cite{Kauffman1984PhysicaD,Kauffman1993book}.  DBNs and DMNs correspond to
$K\propto n$, and SBNs and SMNs correspond to $K\ll n$.  We study
typical dynamical behavior of the random DBNs, SBNs, DMNs, and SMNs.

The number of possible functions generated by a single unit in the
four types of network is compared as follows. The number of all
Boolean functions is equal to $2^{2^n}$, growing doubly exponentially
with $n$. This is equal to the variety of the random mapping on $n$
units.  The number of functions generated by a single unit in DBNs
is large for large $K$ and equal to $2^{2^n}$ when $K=n$.  In
contrast, SBNs, DMNs, and SMNs are limited in terms of the number of
possible functions. The number of all majority vote units is
asymptotically equal to $2^{n^2/2}$; the growth rate is only
exponential.  For the sparse Boolean and majority vote networks, the
number of functions generated by a single unit is equal to $2^{K\log_2
  n}$, growing only algebraically with $n$.  The differences in the
variety of functions in the four cases may result in different
dynamical behaviors of the networks, as we will analyze in the
following.

\section{Distance law in state transitions}

The network state at time $t$ is given in 
vector form as
\begin{equation}
\bm x(t) = \left(x_1(t), \ldots, x_n(t)\right).
\end{equation}
Let $X= \left\{ \bm x\right\}$ be the set of the $N\equiv 2^n$
states. Given a network, the 
state transition is a mapping from $X$ to itself.
We write it briefly as $\bm x(t+1) = f\bm x(t)$.

The dynamics of the distance between two state trajectories has been
studied to characterize dynamics in these networks. We define the normalized Hamming distance
between two states $\bm x$ and $\bm y$ in $X$ by
\begin{equation}
D(\bm x, \bm y) =
\frac{1}{2n} \sum_{i=1}^n \left| x_i-y_i \right|.
\end{equation}
It should be noted that
the distance is restricted to the range $0 \le D(\bm x, \bm y)
\le 1$.
We let $d= D(\bm
x, \bm y)$ and $d^{\prime} = D \left(f\bm x, f\bm y\right)$. The mapping from $d$ to $d^{\prime}$ is a random variable and depends on $\bm x$ and $\bm y$ because $w_{ij}$ is a random variable and many pairs
of $\bm x$ and $\bm y$ realize $d= D(\bm x, \bm y)$. However, we can prove
that $d^{\prime}=\varphi(d)$ for a function $\varphi(d)$ for
any $\bm x$ and $\bm y$ almost always as $n\to\infty$. We call $\varphi(d)$
the distance law.
For DBNs, $\varphi(d) = 0$ ($d=0$) and $\varphi(d)
=1/2$ ($d \ne 0$).  For SBNs \cite{DP1986},
\begin{equation}
\varphi(d) = (1/2)
\left[1-(1-d)^K\right].
\end{equation}
For DMNs \cite{amari1974,AYK1977,Amari1990IEEE},
\begin{equation}
\varphi(d) = (2/\pi) \sin^{-1} \sqrt{d}.
\end{equation}
For SMNs \cite{Derrida1987JPA},
\begin{equation}
\varphi(d) = \sum_{j=0}^K g_{K, j} \left( K \atop j \right)
d^j (1-d)^{K-j},
\end{equation}
where $\left(K \atop j \right)$ is the
binomial coefficient and 
\begin{equation}
g_{K, j} = (2/\pi)\sin^{-1}\sqrt{j/K}.
\end{equation}

The dynamics of the distance under the annealed approximation is given by
$d_{t+1}= \varphi \left(d_t \right)$
\cite{amari1974,DP1986,Derrida1987JPA,kurten1988}.
For all four types of network, $\varphi(0)=0$.  For DBNs,
$\varphi(d)$ is discontinuous at $d=0$, and the dynamics is essentially
the same as that of a random mapping on the $N$ states. Therefore,
various properties of the dynamics such as the number of attractors,
transient length, and cycle period are well characterized
\cite{amari1974,Kauffman1984PhysicaD,Kauffman1993book}.  For SBNs, DMNs, and SMNs,
$\varphi(d)$ is continuous at $d=0$. It is known that $d=0$ is a stable
fixed point of mapping $\varphi$ [i.e., $0<\varphi^{\prime}(0)<1$] only
for SBNs and SMNs with $K= 1$ or 2.  Otherwise, $d_t$
converges to a positive value $\overline{d}$ satisfying
$\overline{d}= \varphi \left(\overline{d}\right)$.  The convergence of the
distance is usually fast and happens after $\sim 10$ steps except for
DBNs where one step is enough for the distance to
converge.

\section{Exponent of state concentration}

The average transient length
before the orbit enters the attractor is
analytically intractable. In addition, it may
not be a good measure of the quickness of the dynamics due to the long-tailed nature of its distribution \cite{BL1996}.
Therefore, we introduce
an alternative order parameter called the
exponent of the state concentration. We
use the so-called state transition graph
\cite{Kauffman1984PhysicaD,Kauffman1993book,Shreim2008NJP}
defined as follows.  A map $f$, either Boolean or majority vote, induces
a graph on $N$ nodes.  Each state $\bm x\in X$ defines a node and has
exactly one outgoing link directed to node $f\bm x$.

Suppose that each of the $N=2^n$ nodes (i.e., states) has a token at
$t=0$. For each $t (\ge 0)$, an application of $f$ moves all the tokens
at each node $\bm x$ to
node $f\bm x$.
Repeated applications of the mapping $f$ elicit concentrations of tokens at specific nodes. We denote by
$f^{-t}\bm x$ the set of nodes whose tokens move to $\bm x$ after $t$
steps and by $\left| f^{-t}\bm x\right|$ the number of tokens at node
$\bm x$ after $t$ steps. Tokens are initially equally distributed, i.e.,
$\left| f^0\bm x\right|=1$ for each $\bm x$, and the total number of
tokens is conserved throughout the dynamics, i.e.,
\begin{equation}
\sum_{\bm x\in X}\left|f^{-t}\bm x\right|=N
\end{equation}
for $t\ge 0$.

\begin{figure}[h]
 \begin{center}
 \includegraphics [height=40mm]{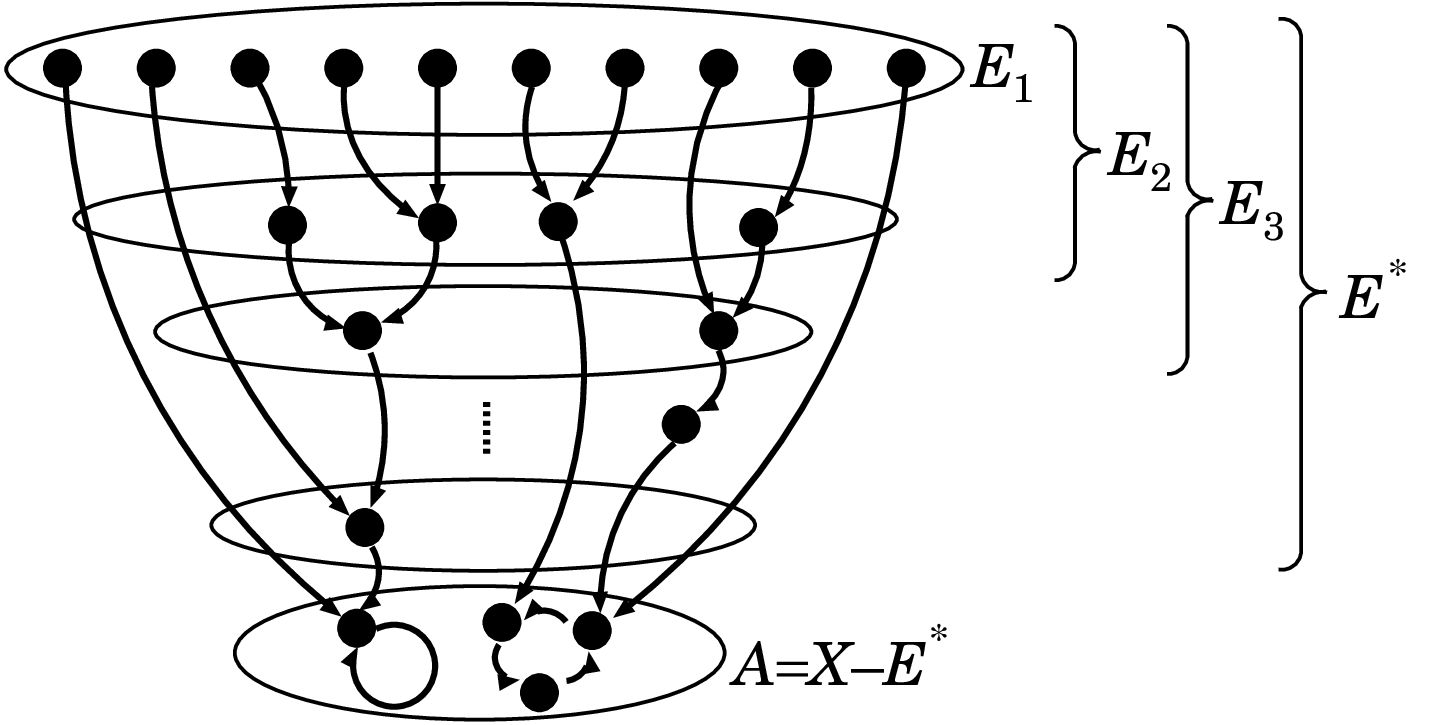}
\caption{Schematic of the dynamics of state concentration.
$E_1$ indicates the Garden of Eden, $E_2$ indicates the nodes that do not have grandparents,
and $A$ indicates the set of attractors.}
\label{fig1}
\end{center}
\end{figure}

The in-degree of node $\bm x$ in the state transition
graph is equal to $\left|f^{-1}\bm x\right|$. The nodes with $f^{-1}\bm
x=\emptyset$, where $\emptyset$ is the empty set, do not have parent
nodes.  The set of such nodes is called the Garden of Eden
\cite{Kauffman1984PhysicaD}
and denoted by $E_1$. The nodes $\bm x\in E_1$
appear only as initial states.  Only the nodes $\bm x\in X-E_1$
retain tokens at $t=1$.  In general, we define $E_t\equiv \left\{ \bm x
  \mid f^{-t}\bm x = \emptyset \right\}$, i.e., the set of nodes
that do not have tokens at time step $t$.  There exists integer
$T$ such that
\begin{equation}
\phi \subset E_1 \subset E_2 \subset \cdots \subset E_T
= E_{T+1} = \cdots \equiv E^{\ast},
\end{equation}
 where
$T$ is the longest transient period and
$E^{\ast}$ is the set of the transient states.
The set of the attractors is given by $A=X-E^{\ast}$
(Fig.~\ref{fig1}).

To quantify the state concentration, we consider
the number of tokens that a token at $\bm x(0)$ and $t=0$ meets after we
  apply $f$.  We write the relationship $f\bm x(0) = \bm x(1)$
succinctly as $\bm x(0)\rightarrow \bm x(1)$.  Let $\overline{S}_1$ be
the conditional expectation $E \left[\left| f^{-1}\bm x(1)\right|
  \mid\bm x(0) \rightarrow \bm x(1)\right]$ of the in-degree of node
$\bm x(1)$ [i.e., the number of parents of $\bm x(1)$] given that
$\bm x(0)\rightarrow \bm x(1)$ and that $\bm x(0)$ is selected with
equal probability (i.e., $1/N$).  In general, we denote by
$\overline{S}_t$ the expected number of $t$-fold ancestor nodes of a
node $\bm x(t)$ conditioned by a state transition path ending at $\bm
x(t)$ through which a token has traveled,
or equivalently, conditioned by the uniformly distributed initial token $\bm x(0)$,
i.e.,
\begin{equation}
\overline{S}_t\equiv
E\left[ \left|f^{-t}\bm x(t)\right| \mid
\bm x(0)\rightarrow \cdots \rightarrow\bm x(t-1)\rightarrow \bm x(t)
\right].
\label{eq:def of S_t}
\end{equation}
Obviously $\overline{S}_0=1$
and the sequence $\{\overline{S}_t\}$ is monotonically nondecreasing
in $t$. If
\begin{equation}
\overline{S}_t = e^{c_t n}, \;  c_t>0,
\end{equation}
holds true for large $n$, the tokens are exponentially concentrated
on nodes having at least a $t$-fold ancestor node.  We refer to
\begin{equation}
c_t= \lim_{n\to \infty}
 \frac{\ln \overline{S}_t}{n}
\label{eq:def of c_t}
\end{equation}
 as the $t$-step exponent of the concentration.
The exponent $c_t$ quantifies the degree of state concentration and is a measure of quickness.
It should be noted that we do not need to explicitly evaluate
the statistics of the transient to calculate $c_t$.

The stochastic symmetry of units makes the calculation of
$\overline{S}_t$ tractable.  To explain the symmetry, we consider the
majority vote network; similar arguments hold true for the Boolean
network. Because the weights $w_{ij}$ are independently and identically
distributed, the probability distribution of the mapping $f$ is invariant
under permutation of $x_1$, $\ldots$, $x_n$, which are passed as the
arguments to $f_i$ ($1\le i\le n$).  In addition, the probability is
invariant under flip of the sign of each $x_i$. Therefore, the following
gauge invariance holds.  First, the probability distribution of $\bm
x(t)$ is invariant under permutation of the unit indices.  Second, the
probability distribution is invariant under a sign flip of any
$x_i(t)$.  Because any state is mapped in a single step to a given $\bm
x$ by permutation and sign flip, all the states are stochastically
equivalent.  Therefore, $\mbox{Prob}\left\{ \left|f^{-1}\bm x\right|=k
\right\}$, for example, is the same for all $\bm x$, and
$\mbox{Prob}\left\{\bm x(0)\rightarrow \bm x(1)\right\}=1/N$ for any
$\bm x(0)$ and $\bm x(1)$.

\begin{figure}[h]
 \begin{center}
 \includegraphics [height=40mm]{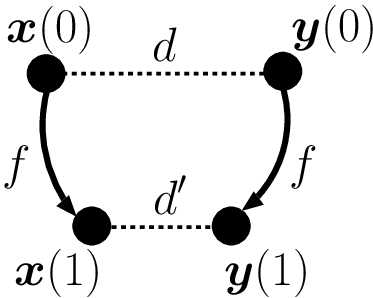}
\caption{Schematic of one-step dynamics of the distance.}
\label{fig:distance transformation}
\end{center}
\end{figure}

We define a conditional probability distribution
\begin{equation}
r_k =
\mbox{Prob} \left\{ \left| f^{-1} \bm x(1)\right|=k \; \mid\bm x(0)
  \rightarrow \bm x(1)\right\},
\end{equation}
which is
the in-degree distribution of node $\bm x(1)$ conditioned by
$\bm x(0)\to\bm x(1)$. The symmetry guarantees that $r_k$ is independent of $\bm x(0)$ and $\bm x(1)$.
Let us compute
\begin{equation}
\overline{S}_1 = \sum_{k=1}^N k r_k.
\end{equation}
We denote by
$\bm y(0)$ a node such that
$D \left(\bm x(0), \bm y(0)\right) = d$ for a given $\bm x(0)$.
The number of nodes with distance $d$ away from $\bm x(0)$ is given by
\begin{equation}
\left(n \atop nd
\right) \approx e^{nH(d)},
\end{equation}
where
\begin{equation}
H(d)\equiv -d \ln d - (1-d) \ln (1-d)
\end{equation}
is the entropy.
Because $\varphi(d)$ is the probability that the $i$th components of 
$f\bm x(0)$ and $f\bm y(0)$
differ for any $i$, 
the probability that
$D \left(f\bm x(0), f\bm y(0)\right)=d^{\prime}$
(see Fig.~\ref{fig:distance transformation} for a schematic illustration of this situation) is given by the binomial distribution as follows:
\begin{equation}
\psi\left(d^{\prime} \mid d \right) \equiv
 \left( n \atop {nd^{\prime}} \right) \varphi(d)^{nd^{\prime}}
 \left[1-\varphi(d)\right]^{n \left(1-d^{\prime}\right)}.
\end{equation}
In particular,
\begin{equation}
\psi(0\mid d) = \left[1-\varphi(d)\right]^n =
\mbox{Prob} \left\{ f \bm y(0) = \bm x(1)
\mid \bm x(0)\rightarrow \bm x(1)\right\}.
\end{equation}
By using the saddle-point
approximation, we obtain
\begin{align}
\overline{S}_1 =& \sum^n_{nd=0} \left(n \atop
{nd}\right) \psi(0\mid d)\notag\\
\approx& \int \exp 
 n \left\{ H(z) + \ln \left[1-\varphi(z)\right] \right\} dz\propto e^{nc_1},
  \label{eq:amInt}
\end{align}
where
\begin{equation}
c_1 = H
 \left(d^* \right)+ \ln \left[1-\varphi \left(d^* \right)\right]
\end{equation}
and
\begin{equation}
d^* = \mathop{\arg\max}_{d} \left\{ H(d)+ \ln
 \left[1-\varphi(d)\right]\right\}.
\end{equation}

To evaluate $c_t$ in general, we consider a $t$-step state transition
path $X_t = \{\bm x(0)\rightarrow \bm x(1)\rightarrow\cdots \rightarrow
\bm x(t) = \bm x^*\}$ ending at $\bm x^*$ and calculate the conditional
probability that another path $Y_t = \{\bm
y(0)\rightarrow\cdots\rightarrow\bm y(t)\}$ ends at the same $\bm x^*$.
$\overline{S}_t$ is the expectation of the number of such $t$-step
paths.  Let us denote the distance $D(\bm x(t^{\prime}), \bm
y(t^{\prime}))$ by $d_{t^{\prime}}$, where $0\le t^{\prime}\le t$ and
$d_t = 0$.  Then, under the Markov assumption, the probability of path
$Y_t$ conditioned by path $X_t$ is represented in terms of the distances
of the two sequences, i.e., $d_{t^{\prime}}, 0 \le t^{\prime} \le t$,
by
\begin{equation}
\mbox{Prob} \{Y_t\mid X_t\} = 
\prod_{t^{\prime}=1}^{t^{\prime}=t}
\psi(d_{t^{\prime}}\mid d_{t^{\prime}-1}).
\label{eq:Markov assumption}
\end{equation}
The Markov assumption is valid for
large $n$ and a finite $t$.  
Because there are 
$\left( n \atop {nd_0} \right)$
states $\bm y(0)$ possessing distance $d_0$ from
$\bm x(0)$, the expected number of paths is given by the 
integration of $\left( n \atop {nd_0} \right)
\mbox{Prob }\{Y_t\mid X_t\}$ over all the possible distance
sequences $\{d_0, \ldots , d_{t-1}, d_t=0\}$.  By using the saddle-point approximation,
we have
\begin{equation}
\overline{S}_t = \exp
\left[ n H(d_0^*) + \sum_{t^{\prime}=1}^t
\ln \psi(d_{t^{\prime}}^*\mid d_{t^{\prime}-1}^*)
 \right]
\label{eq:S_t}
\end{equation}
for large $n$, where $d_{t^{\prime}}^*$ are the maximizers of the integrand in the path
integration. Equation~\eqref{eq:S_t} implies
\begin{align}
c_t =& \sum_{t^{\prime}=1}^t
\left\{ H(d_{t^{\prime}-1}^*) + d_{t^{\prime}}^*\ln \varphi(d_{t^{\prime}-1}^*)
\right.\notag\\
&\left. +(1-d_{t^{\prime}}^*)\ln \left[1-\varphi(d_{t^{\prime}-1}^*)\right]
\right\}.
\label{eq:general c_t}
\end{align}

For example, for $t=2$, we obtain
\begin{align}
c_2 =&
H(d_0^*) + H(d_1^*)  + \ln \left[1-\varphi
  \left(d_1^*\right)\right] + d_1^* 
\ln \varphi \left( d_0^*\right)\notag\\
 +& \left( 1-d_1^*\right) \ln \left[1-\varphi
 \left(d_0^*\right)\right],
\end{align}
where
\begin{align}
\{d_0^*, d_1^*\}=&
\arg \max_{d_0,d_1}\left\{H(d_0) + H(d_1)  + \ln
\left[1-\varphi\left(d_1\right)\right]\right.
\notag\\
& + \left.d_1\ln \varphi \left(
d_0\right) + \left( 1- d_1\right) \ln \left[1-\varphi
  \left(d_0\right)\right] \right\}.
\end{align}

\begin{figure}[h]
 \begin{center}
\rotatebox{-90}{\includegraphics [keepaspectratio=true,height=80mm]{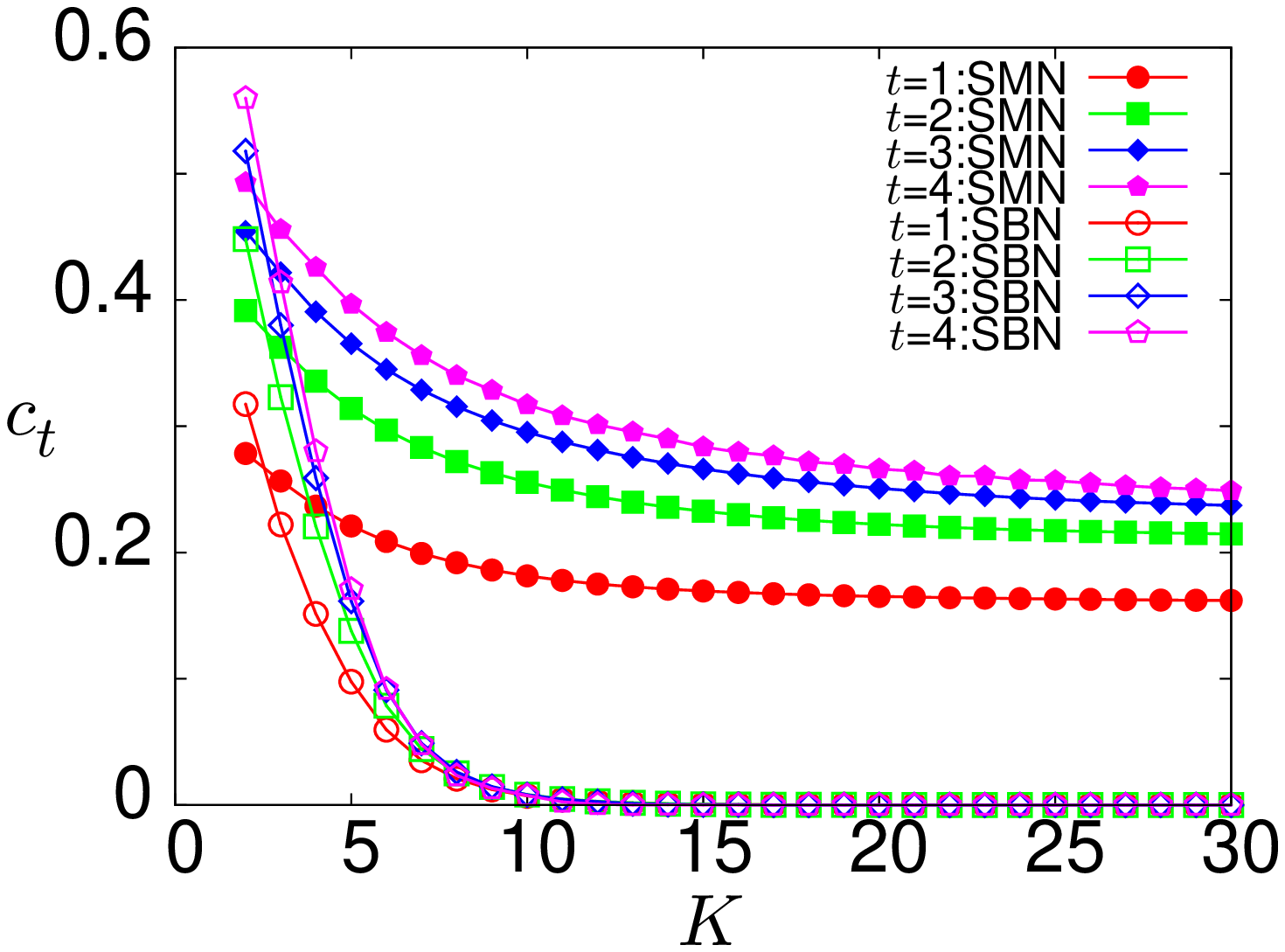}}
  \caption{(Color online) $c_1$, $c_2$, $c_3$, and $c_4$
for SMNs and SBNs.}
\label{fig:c_ell vs K}
 \end{center}
\end{figure}

On the basis of the expression of $\varphi$ for SBNs and SMNs shown
before, the dependence of $c_1$, $c_2$, $c_3$, and $c_4$
on $K$ is plotted in
Fig.~\ref{fig:c_ell vs K}.  For SBNs, $c_t$ ($1\le t\le 4$)
converges to $0$ quickly as $K$ increases.  For DBNs, which is the
case for $K=n$, we trivially obtain $c_t=0$ at least for small $t$
because $f$ is equivalent to the random mapping.  Figure~\ref{fig:c_ell
vs K} indicates that the state concentration occurs only for
very small $K$ in the random Boolean network.  In contrast, the
state concentration occurs even for large $K$ in majority vote
networks. In particular, for DMNs with $K\to n$, we obtain
$c_1\approx 0.157$ \cite{Amari1990IEEE}.
Figure~\ref{fig:c_ell vs K} also indicates that
the state concentration quickly proceeds as $t$ increases, except in DBNs.

\begin{figure}[h]
 \begin{center}
 \rotatebox{-90}{\includegraphics [keepaspectratio=true,height=80mm]{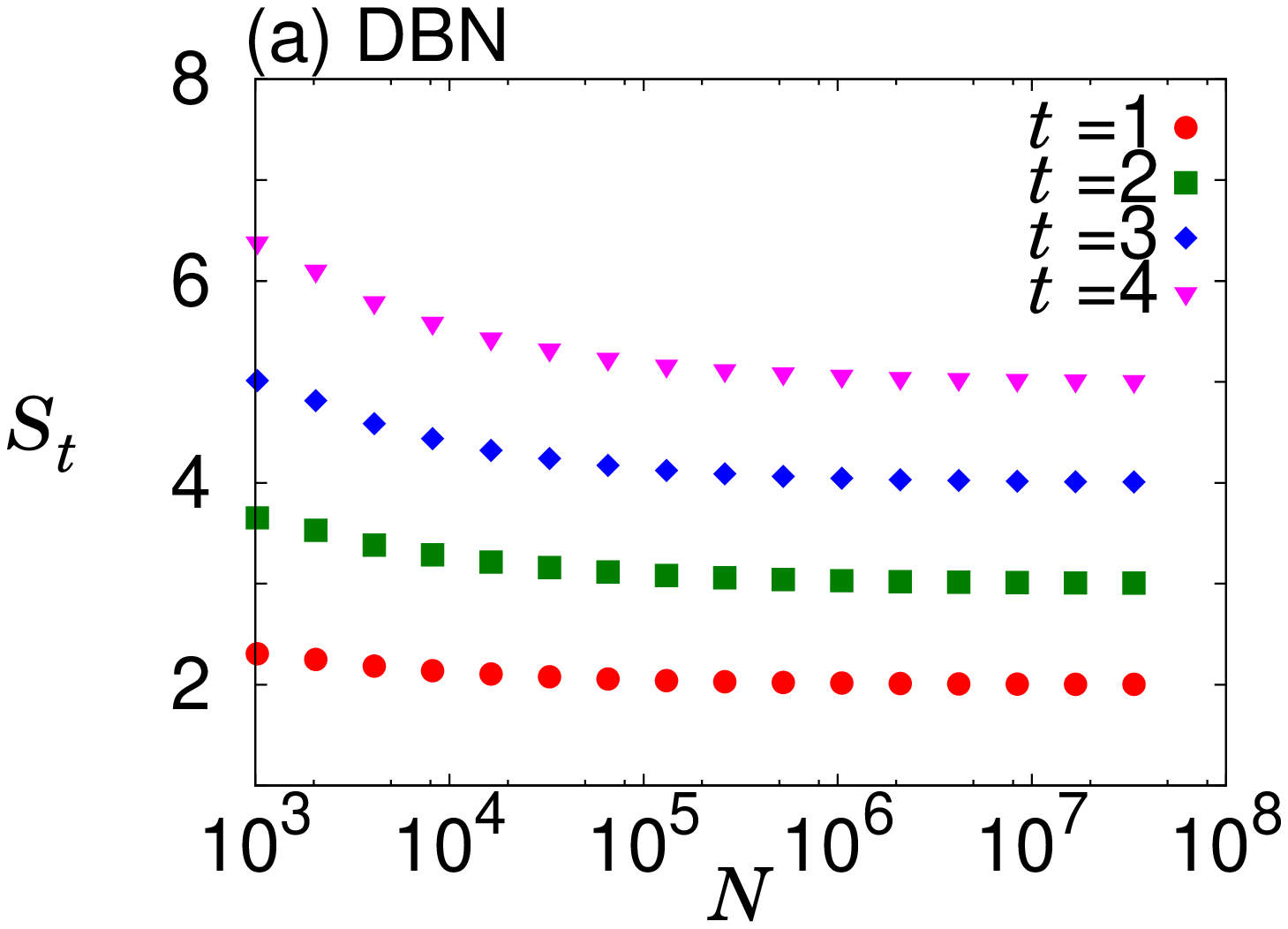}}
 \rotatebox{-90}{\includegraphics [keepaspectratio=true,height=80mm]{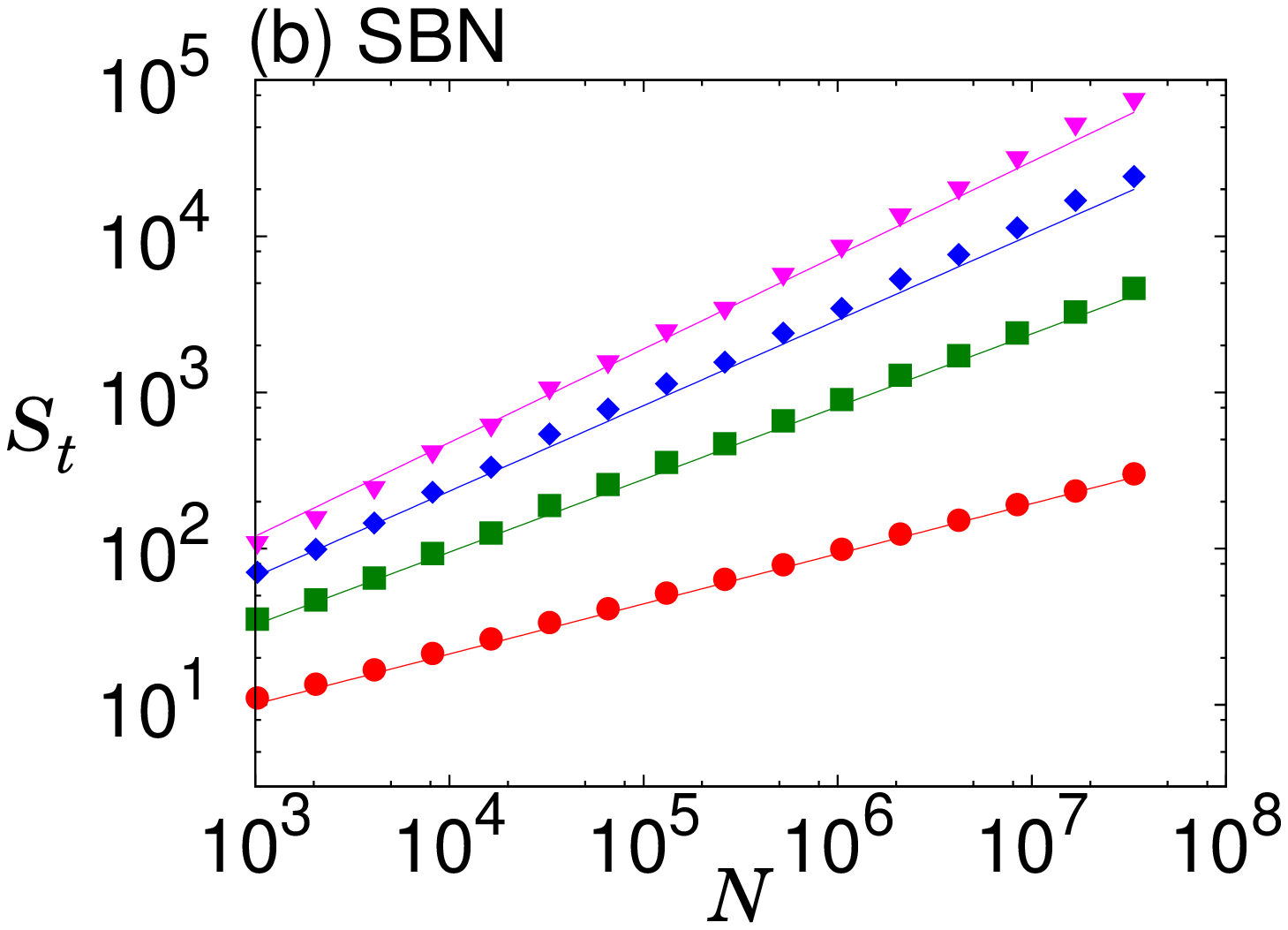}}
 \rotatebox{-90}{\includegraphics [keepaspectratio=true,height=80mm]{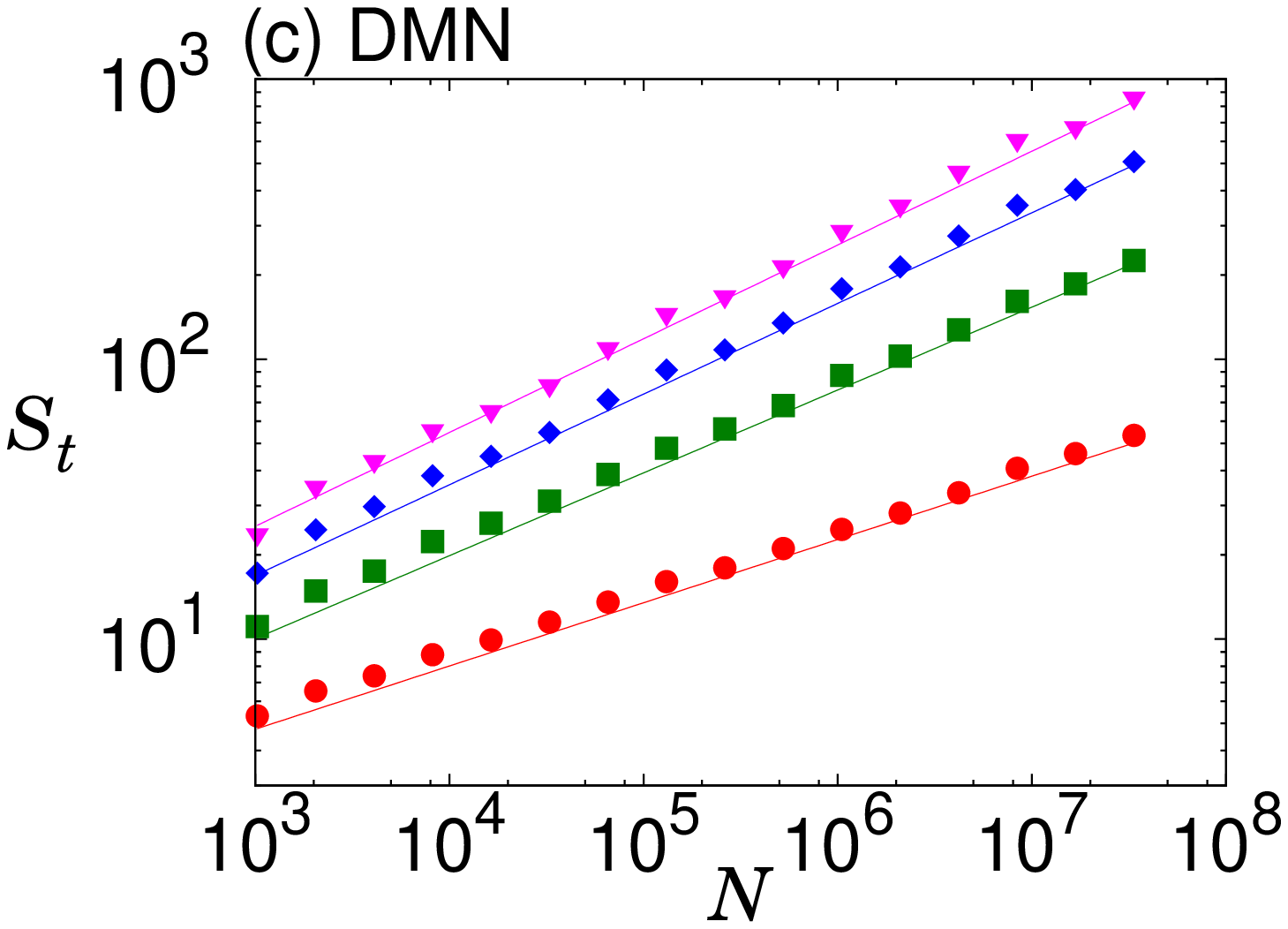}}
 \rotatebox{-90}{\includegraphics [keepaspectratio=true,height=80mm]{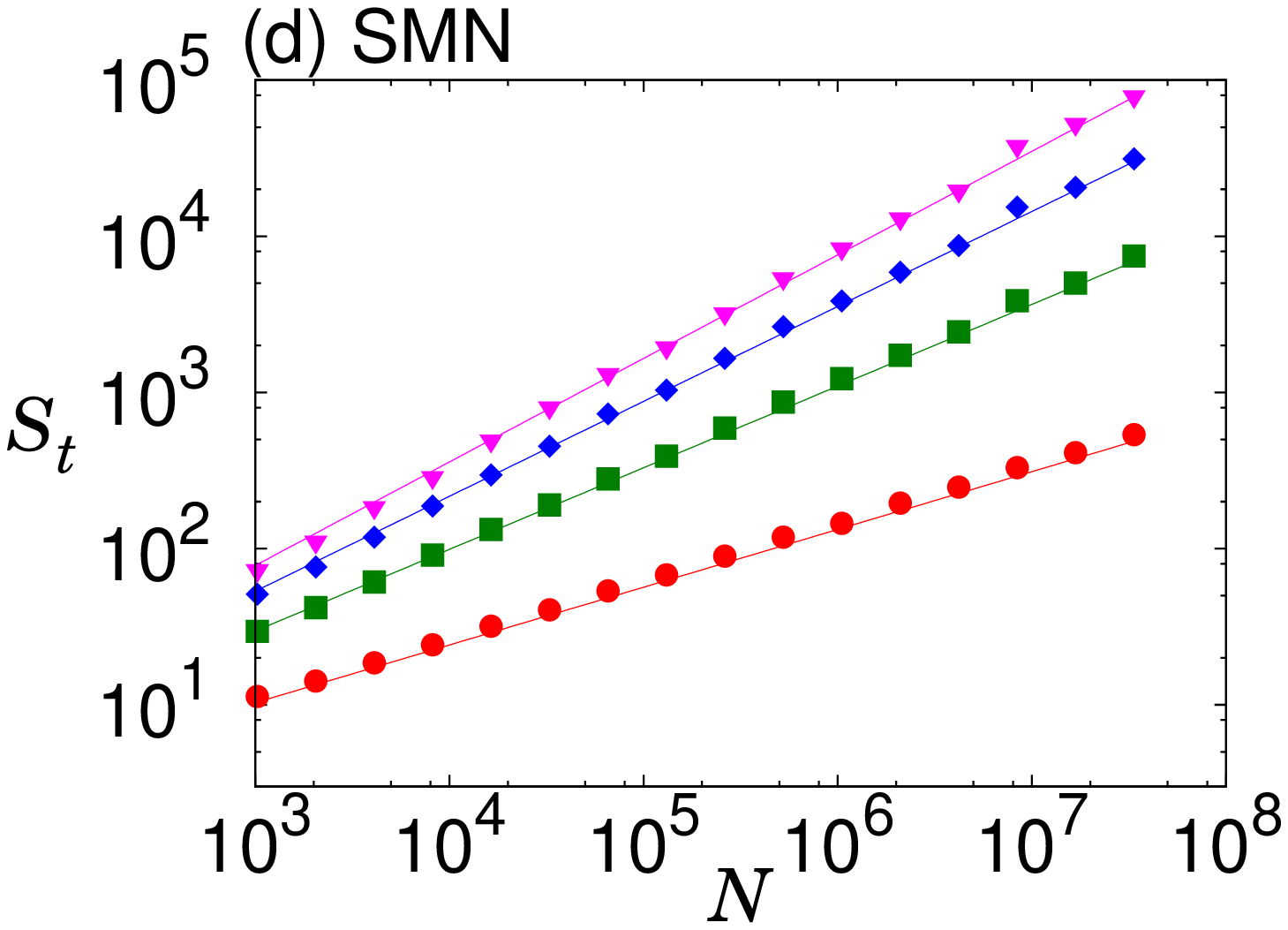}}
 \caption{(Color online) $\overline{S}_t$ exponentially increases with $n$ except in DBNs. (a) DBNs, (b) SBNs with $K=3$,
(c) DMNs, and (d) SMNs with $K=3$. The lines in (b), (c), and (d) indicate the theoretical estimates (see Table~\ref{tab:S_t}).
Each point in the figure represents the average of $\overline{S}_t$
over $10^3$ realizations of the network.}
\label{fig:S_t compare}
 \end{center}
\end{figure}

We verified Eq.~\eqref{eq:def of c_t}
by comparing
$\overline{S}_t$ obtained from direct numerical simulations (i.e., 
Eq.~\eqref{eq:def of S_t}) and
$c_t$ generally given by
Eq.~\eqref{eq:general c_t}.
The results shown in Fig.~\ref{fig:S_t compare} indicate that
the theory (lines) seems to agree
with numerical results at least for large $n$;
although the largest $n$ value shown in the figure is
only $n = 25$.
Therefore, the Markov assumption (Eq.~\eqref{eq:Markov assumption}) implicitly assumed for $t=2$, 3, and 4 in drawing Fig.~\ref{fig:c_ell vs K}
roughly holds true at least up to $t\approx 4$ for large $n$.

\begin{table}
\begin{center}
\caption{Theoretical estimates of $c_t=\ln\overline{S}_t/n$.}
\begin{tabular}{|c|ccc|}\hline
$t$ & SBNs & DMNs & SMNs\\ \hline
1 & 0.223 & 0.157 & 0.256\\
2 & 0.323 & 0.205 & 0.363\\
3 & 0.380 & 0.225 & 0.421\\
4 & 0.416 & 0.232 & 0.460\\ \hline
\end{tabular}
\label{tab:S_t}
\end{center}
\end{table}

\begin{figure}[h]
 \begin{center}
\includegraphics [keepaspectratio=true,height=60mm]{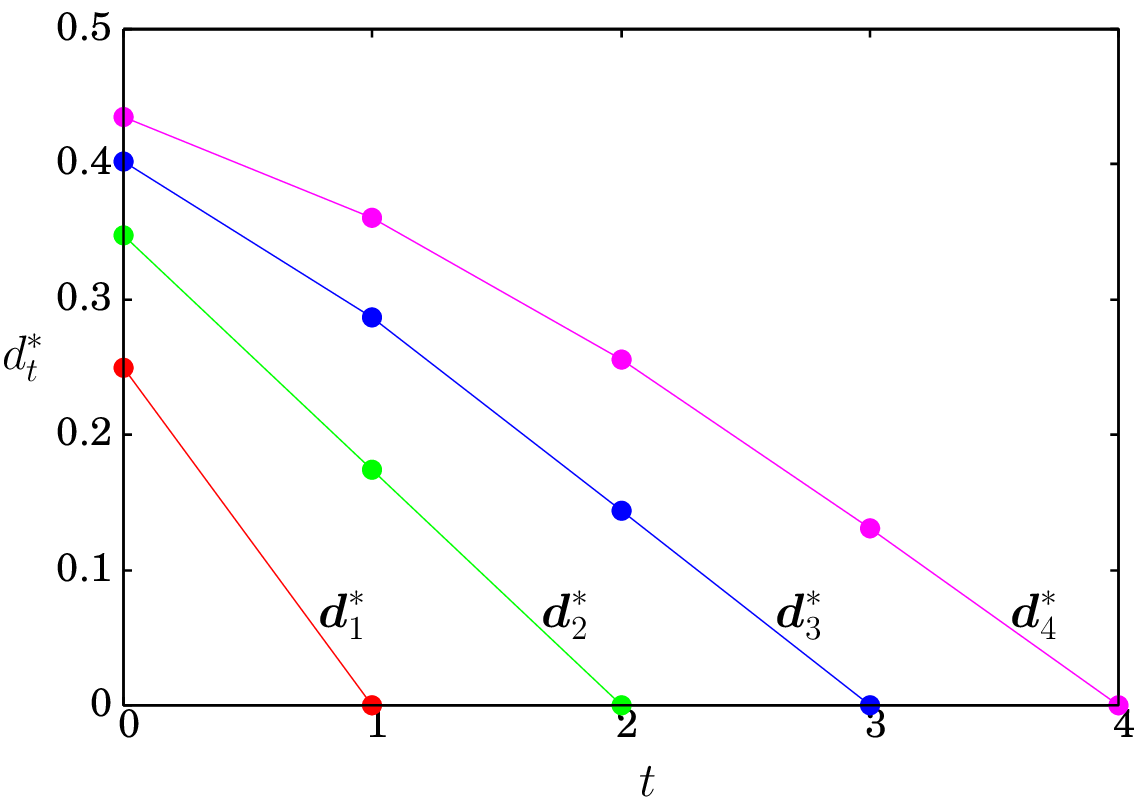}
  \caption{(Color online) Sequence of the most likely distances
$\bm {d}^{\ast}_t = \left\{d^{\ast}_0, d^{\ast}_1, \cdots,
d^{\ast}_{t-1}, d^{\ast}_t=0 \right\}$ for $1\le t\le 4$.}
\label{fig:d*}
 \end{center}
\end{figure}

Theoretically,
most sequences $Y_t$ that meet $X_t$ after $t$
steps of state transition own the sequence of distance given by
${\bm{d}}^{\ast}_t = \left\{ d^{\ast}_0, d^{\ast}_1, \cdots,
d^{\ast}_{t-1}, d^{\ast}_t=0 \right\}$.
In particular, a
majority of the initial states $Y_0$ is
initially separated from $X_0$ by $d^{\ast}_0(t)$.
Figure~\ref{fig:S_t compare} suggests that this is the case at least
up to $t\approx 4$ for large $n$. 
The sequence of distances $\bm{d}^{\ast}_t$ is shown 
for $1\le t\le 4$ in
Fig.~\ref{fig:d*}.

\section{In-degree distribution of the state transition graph}

We calculate the incoming degree distribution
\begin{equation}
p_k = \mbox{Prob} \left\{
\left|f^{-1}\bm x\right|=k\right\},
\end{equation}
 where $k$ is the in-degree of a state and
$\sum_{k=0}^N p_k=1$.
Because each node in the state transition graph has exactly one outgoing link,
 we have
\begin{equation}
\left<k\right>=\sum_{k=0}^N k p_k=1,
\end{equation}
where $\left<\cdot\right>$ indicates the expectation.
Because $r_k = k p_k$
\cite{amari1974}
(also see \cite{NSW2001} for an example),
we obtain
\begin{equation}
\left<k^2\right>= \sum_{k=0}^N k^2 p_k = 
\sum_{k=0}^N k r_k = \overline{S}_1=e^{nc_1}.
\end{equation}
Therefore, $c_1>0$ indicates that $\left<k^2\right>$ diverges in the
limit of $N=2^n\to\infty$, reminiscent of the scale-free property of the
state transition graph \cite{NSW2001,Newman2005comp,dorogovtsev08rmp}.

\begin{figure}[h]
 \begin{center}
\rotatebox{-90}{\includegraphics [keepaspectratio=true,height=80mm]{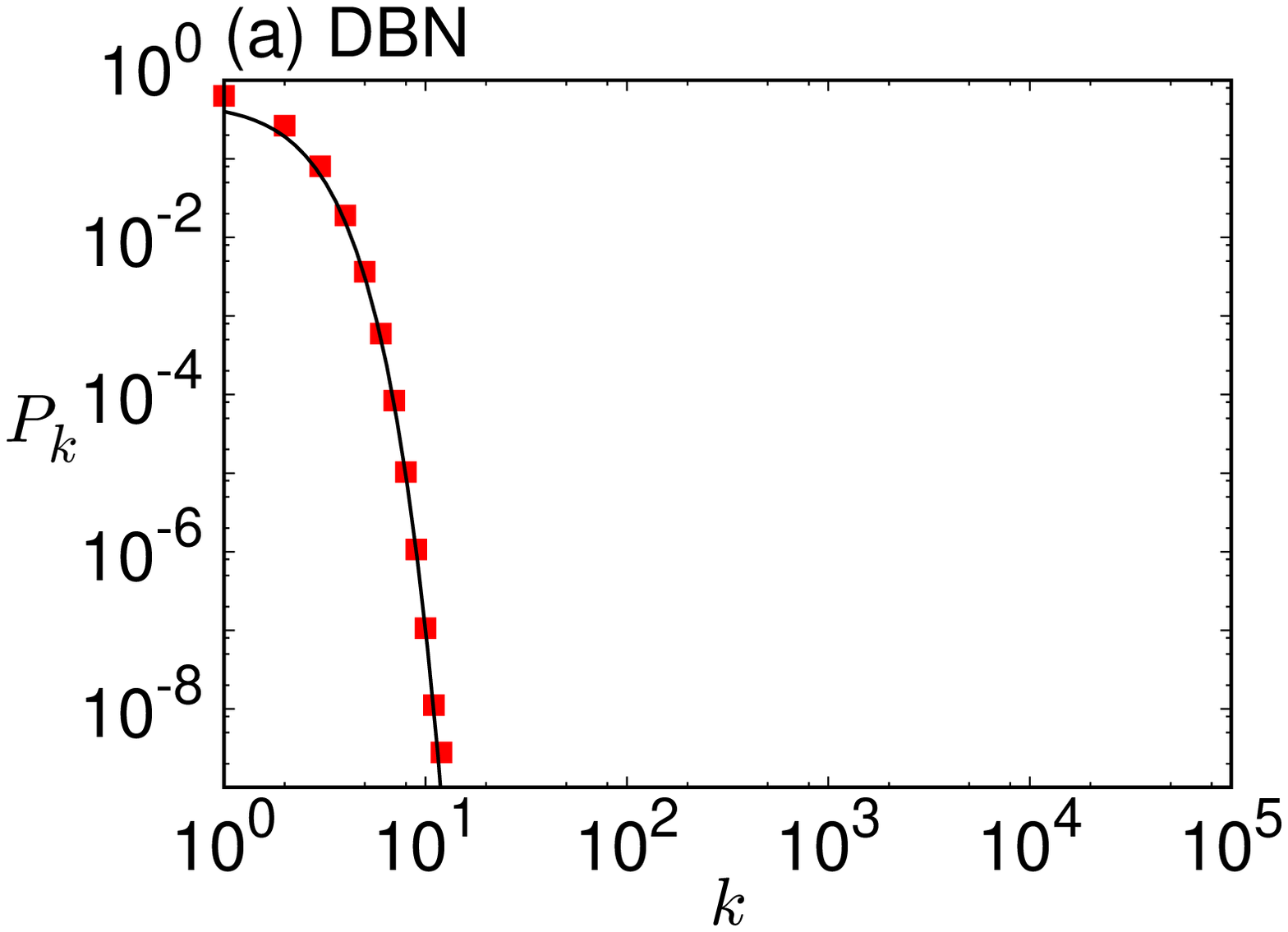}}
\rotatebox{-90}{\includegraphics [keepaspectratio=true,height=80mm]{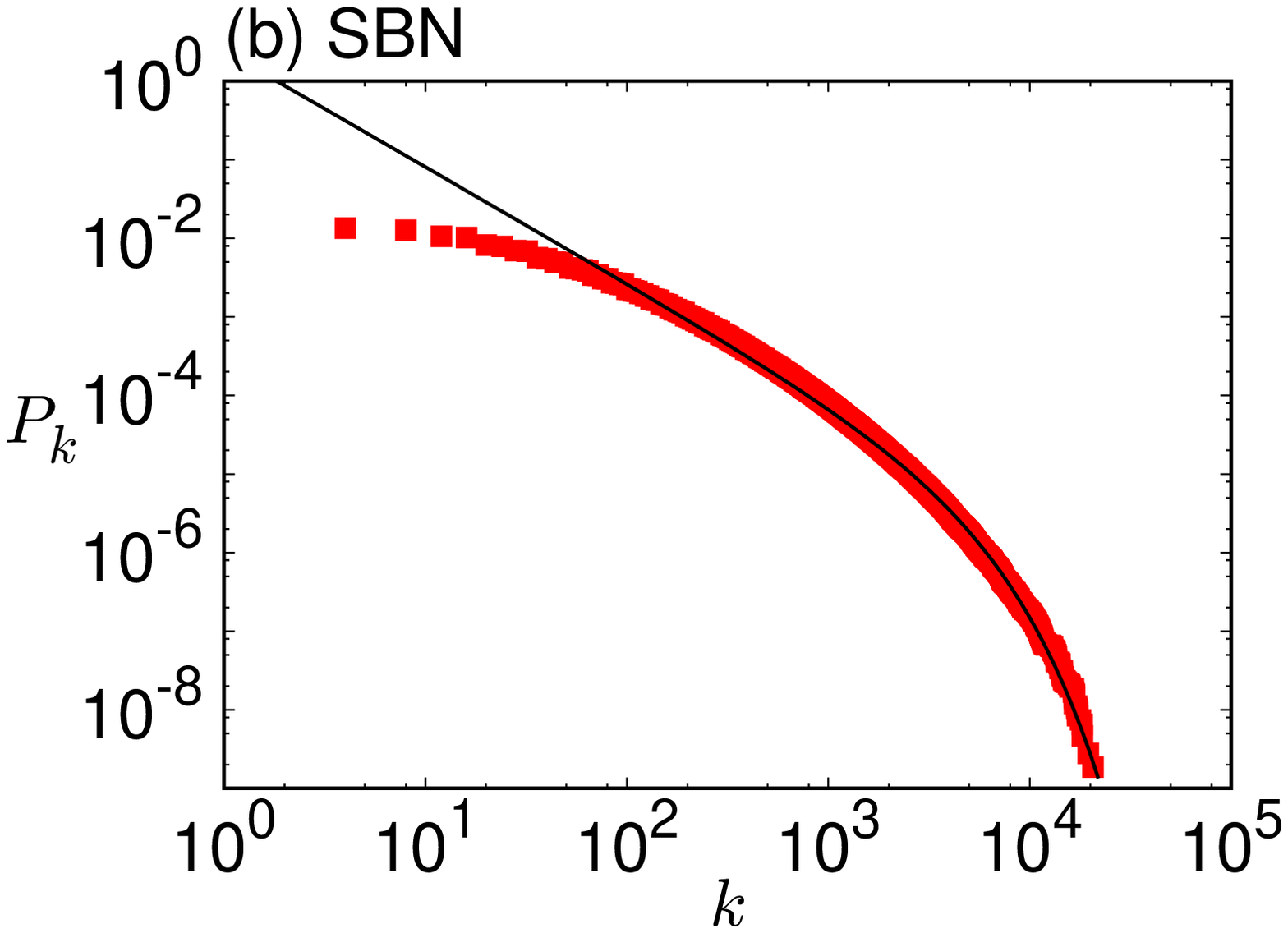}}
\rotatebox{-90}{\includegraphics [keepaspectratio=true,height=80mm]{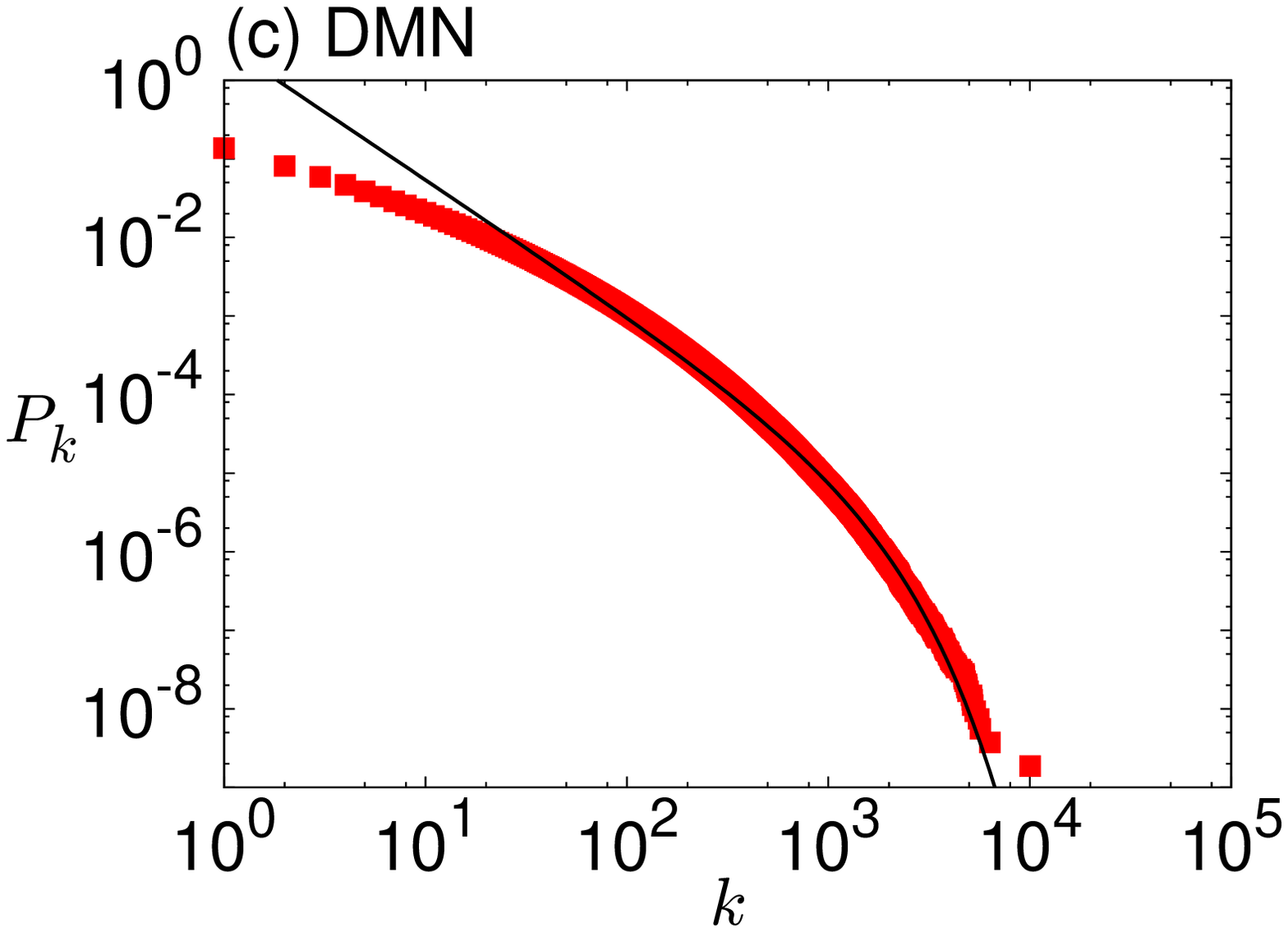}}
\rotatebox{-90}{\includegraphics [keepaspectratio=true,height=80mm]{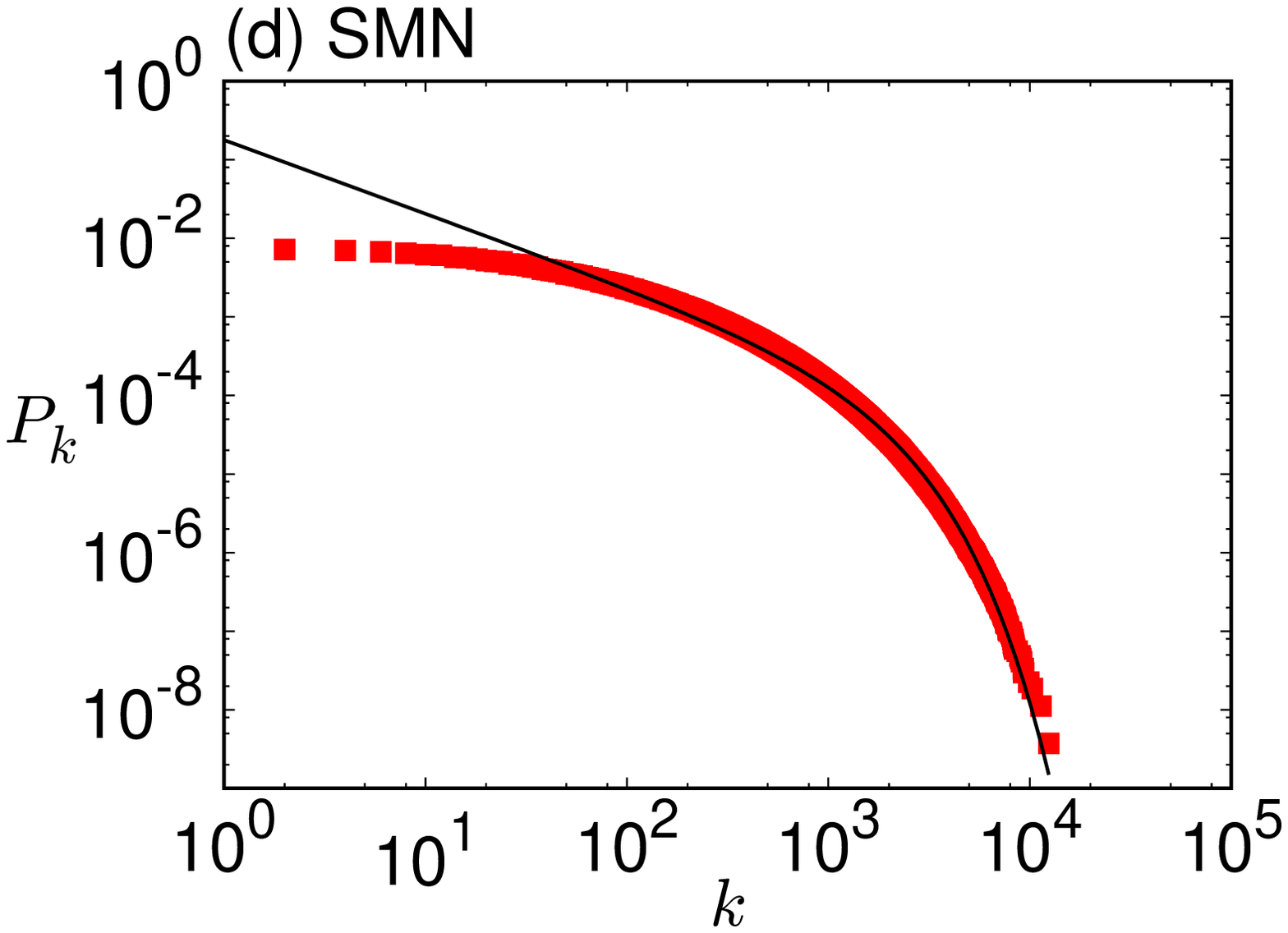}}
 \caption{(Color online) Complementary in-degree distribution of $p_k$ (i.e.,
$P_k\equiv \sum_{k^{\prime}=k}^N p_{k^{\prime}}$) of the state transition 
graph. (a) DBNs ($n=30$ and $K=30$). The numerical results are shown by squares, and the Poisson distribution with mean 1 is indicated by
the line. (b) SBNs ($n=30$ and $K=3$).
(c) DMNs ($n=30$ and $K=30$).
(d) SMNs ($n=30$ and $K=3$).
The fitting curves,
$P_k\propto k^{-1.48} \exp \left\{-0.000298 k \right\}$,
$P_k\propto k^{-1.72} \exp \left\{-0.000985 k \right\}$,
and $P_k\propto k^{-0.936} \exp \left\{-0.000786 k \right\}$,
for (b), (c), and (d), respectively,
obtained from the least square error method, are shown by the lines as guides to
the eyes.}
\label{fig:p_k}
 \end{center}
\end{figure}

For DBNs, the state transition graph is the directed random
graph in which $p_k$ obeys the Poisson
distribution (i.e.,
$p_k = 1/e k!$) with mean and variance 1 \cite{Kauffman1984PhysicaD,Kauffman1993book}. Therefore, $\left<k^2\right>=2$, proving that
$c_1=0$ for DBNs (i.e., there is
no exponential state concentration). This is consistent with
Fig.~\ref{fig:S_t compare}(a) (circles).
Figure~\ref{fig:c_ell vs K} suggests that
$c_1\approx 0$ when $K$ is larger than $\cong 10$.
Therefore,
the in-degree distribution of the state transition graph is also narrow
for SBNs with $K\ge 10$.
We verified that the numerically obtained in-degree distribution
for the random Boolean network with $n=30$ and $K=20$ approximately obeys the Poisson distribution [Fig.~\ref{fig:p_k}(a)].

In contrast, the positive value of $c_1$ found for SBNs with small $K$, DMNs, and SMNs (Fig.~\ref{fig:c_ell vs K})
indicates that $\left<k^2\right> (=\overline{S}_1)$ 
diverges exponentially in $n$.
This is actually the case, as shown in
Figs.~\ref{fig:S_t compare}(b)--\ref{fig:S_t compare}(d).
For scale-free networks with $p_k \approx k^{-\gamma}$,
the extremal criterion would lead to
$\gamma\approx (c_1+3\ln 2)/(c_1+\ln 2)$
\cite{Newman2005comp,dorogovtsev08rmp}.
However, the in-degree distribution numerically obtained for majority vote networks,
shown in Figs.~\ref{fig:p_k}(c) and ~\ref{fig:p_k}(d),
 deviates from a power law. The in-degree distribution
numerically obtained for the SBNs (Fig.~\ref{fig:p_k}(b)) 
is also different from a power law
\cite{Shreim2008NJP}.
To guide the eyes,
fitting curves on the basis of a power law
with an exponential cutoff are shown by the lines
in Figs.~\ref{fig:p_k}(b)--\ref{fig:p_k}(d).
In fact, the power law is not the
only distribution that yields the divergence of
$\left<k^2\right>$.
In the present case, the position of the exponential
cutoff may mildly diverge as $n$ becomes large.

\section{Discussion and conclusions}

In summary, we provided a unified framework for analyzing the state
concentration. We found that exponential
state concentration occurs in
SBNs with small $K$ and the majority vote networks with arbitrary $K$ (i.e., the DMNs and SMNs), but not in DBNs.
We also revealed the long-tailed distributions
of the in-degree of the state transition graph in SBNs, DMNs, and SMNs,
but not in DBNs.

We briefly discuss the relationship between the quickness,
measured by the exponent of the state concentration in this study, and the
robustness of the dynamics.  The robustness of the dynamics is often
measured in terms of damage spreading. It is a long-term property
concerning the stability of $d=0$ for mapping $\varphi$. As we
mentioned, $d=0$ is an unstable fixed point of $\varphi$ unless $K=1$ or
2. Although SMNs with $K=1$ or 2 satisfies quickness and robustness,
we do not discuss these cases because the dynamics in these cases is
just frozen
\cite{kurten1988,Derrida1987JPA}.  Here we consider a simpler measure of
robustness based on a one-step property, i.e., how a difference by a
single bit evolves after a single application of $f$.  This is
essentially the same as the Boolean derivative, a measure of the
robustness used for analyzing random Boolean networks
\cite{Luque1997PRE,Luque2000PhysicaA,Kauffman2004PNAS,Shmulevich2004PRL}.  In SMNs, the probability that a single
bit flip in input
results in a bit flip after the application of $f$ is
given by 
\begin{equation}
g_{K, 1}=(2/\pi)\sin^{-1}\sqrt{1/K}.
\end{equation}
Because $g_{K, 1}$ is equal to $0.5$ for $K=2$ and decreases according
to $\approx 2/(\pi\sqrt{K})$ as $K$ increases, the random majority vote
network is robust except for very small $K$.  However, in the random
Boolean network, the same flip probability is equal to $1/2 -
1/2^{2^K}$, which quickly approaches $1/2$ as $K$ increases.  In
particular, the flip probability for the Boolean network at $K=2$ is
equal to $7/16=0.4375$, which is close to the values for the majority vote
network with $K=2$ (i.e., 0.5) and $K=3$ (i.e., 0.392). Although the
Boolean network with $K=1$ has sufficient robustness, the dynamics in
this case is frozen \cite{Flyvbjerg1988JPA}.  Therefore, we consider that
the robustness of the one-step dynamics holds true in DMNs, but not
in DBNs, SBNs, and SMNs. By combining this observation with our
main results, we consider that quickness and robustness are suitably
balanced in DMNs, but not in DBNs, SBNs, and SMNs.

\end{document}